\documentclass[fleqn,usenatbib]{mnras}

\usepackage{newtxtext,newtxmath}
\usepackage[T1]{fontenc}

\DeclareRobustCommand{\VAN}[3]{#2}
\let\VANthebibliography\thebibliography
\def\thebibliography{\DeclareRobustCommand{\VAN}[3]{##3}\VANthebibliography}

\usepackage{graphicx}	% Including figure files
\usepackage{amsmath}	% Advanced maths commands
\title[Infrared lags in AGN]{Infrared lags in the light curves of AGN measured using a deep survey}

\author[E. Elmer et al.]{
E. Elmer,$^{1}$\thanks{E-mail: elizabeth.elmer@nottingham.ac.uk}
M. Merrifield,$^{1}$
O. Almaini,$^{1}$
W. G. Hartley,$^{2}$
D. T. Maltby $^{1}$
\\
$^{1}$School of Physics and Astronomy, University of Nottingham, University Park, Nottingham, NG7 2RD, UK\\
$^{2}$Department of Astronomy, University of Geneva, CH-1205 Versoix, Switzerland\\
}

\date{Accepted XXX. Received YYY; in original form ZZZ}

\pubyear{2020}

\begin{document}
\label{firstpage}
\pagerange{\pageref{firstpage}--\pageref{lastpage}}
\maketitle

\begin{abstract}
    Information on the structure around active galactic nuclei (AGN) has long been derived from measuring lags in their varying light output at different wavelengths.  In principle, infrared data would reach to larger radii, potentially even probing reprocessed radiation in any surrounding dusty torus.  In practice, this has proved challenging because high quality data are required to detect such variability, and the observations must stretch over a long period to probe the likely month-scale lags in variability. In addition, large numbers of sources would need to be observed to start searching for any patterns in such lags.  Here, we show that the UKIDSS Ultra Deep Survey, built up from repeated observations over almost a decade, provides an ideal data set for such a study.  For 94 sources identified as strongly-varying AGN within its square-degree field, we find that the $K$-band light curves systematically lag the $J$-band light curves by an average of around a month. The lags become smaller at higher redshift, consistent with the band shift to optical rest-frame emission.  The less luminous AGN also display shorter lags, as would be expected if their physical size scales with luminosity.  
\end{abstract}


%%%%%%%%%%%%%%%%%%%%%%%%%%%%%%%%%%%%%%%%%%%%%%%%%%

%%%%%%%%%%%%%%%%% BODY OF PAPER %%%%%%%%%%%%%%%%%%

\section{Introduction}
    \label{sec:Intro}
    
    The central regions of most Active Galactic Nuclei (AGN) are almost impossible to resolve with current observational techniques, so less direct measures have been used to infer information about their structure. In particular, if the emission from the AGN is varying significantly, then delays or "lags" in the variability between different wavebands can be interpreted as arising from the reprocessing of energy emitted near the centre of the AGN, with, in the simplest picture, the lag representing the extra light travel time for the reprocessed emission \citep[e.g.][]{Sergeev2005, Cackett2007}.  If the environment becomes cooler at larger distances from the AGN, one might expect the lags to become larger when emission is reprocessed to longer wavelengths, as indeed has been observed in quite a number of low-redshift AGN.  For example, \citet{Arevalo2008, Arevalo2009} investigate lags between the X-ray and optical light curves, while \citet{Sergeev2005} and \citet{Edelson2015} examine optical lags relative to the ultraviolet, uncovering clear evidence for time delays on timescales ranging from hours to days. 
    
    It would be very interesting to push such analyses to longer wavelengths, as one might expect any reprocessing in the near-infrared (NIR) to be occurring in the putative outer dusty torus that perhaps joins on to the inner accretion disk \citep{Koshida2009}.  Indeed, work that has been done at these wavelengths suggests longer lags between optical and NIR, though the difficulty of such observations and the heterogeneity of the small number of sources observed has yielded lags ranging from a few days \cite[e.g.][]{Lira2015, Shappee2014} to months \cite[e.g.][]{Lira2011, Oknyansky2014, Koshida2009}. Ideally, one would want a much larger sample of AGN observed homogeneously and at high quality over years in both the infrared and the optical, in order to study this phenomenon systematically across the population.  
    
    Fortunately, such a data set exists.  The United Kingdom Infrared Telescope (UKIRT) Infrared Deep Sky Survey (UKIDSS) Ultra Deep Survey (UDS), the deepest NIR survey over $1\ \mathrm{deg^{2}}$ of sky \cite[see][and Almaini et al. in prep]{Lawrence2007}, was built up from repeated observations spread over almost a decade , and the relatively bright AGN can have their photometry measured from individual observations, allowing their variability to be monitored over this whole period. The survey obtained data in both $J$ and $K$ infrared bands, and over a wide range in redshift ($0.2<z<1.2$) these bands correspond to rest-frame optical ($\lambda < 1\mu$m) and near-infrared emission respectively,  so comparing the two bands would allow any lag between these wavebands to be measured.
    
    To this end, we have extracted photometry for a sample of variable AGN from the $J$- and $K$-band UDS DR11 observations. In Section~\ref{sec:Method} we describe the sample selection and the cross-correlation method used to determine lags. In Section~\ref{sec:Results}, we present the detection of a systematic lag in these data, and how it depends on AGN redshift and luminosity.  The implications of these measurements are discussed in Section~\ref{sec:Discussion}.

\section{Data and Analysis}
    \label{sec:Method}
    
    The AGN used in this study were selected based on their variability, using the method described in \citet{Elmer2020}, where a source is considered variable based on a $\chi^{2}$ comparison to a non-varying light curve. In the current work we extend the method to both the $J$- and $K$-band data, co-adding images into deep stacks corresponding to eight years of observations. This analysis yielded an initial total of 595 variable sources, 177 of which were identified as variable in both bands. However, this binning is too crude to search for variability on shorter timescales, so we also produced photometry for the AGN based on month-long snapshots.  The lower signal-to-noise level of these shorter observations meant that the fainter sources were too noisy for the subsequent analysis, so we imposed a more stringent variability cut on the \cite{Elmer2020} analysis of $\chi^{2} > 100$, leaving a total of 94 variable sources.  We have spectroscopic redshifts for 72 of these sources, and used photometric redshifts for the remaining 22 AGN \citep[see][]{Elmer2020}. We note, however, that excluding the objects with photometric redshifts has no significant impact on the results presented below.
    
    The UDS data were taken between 2005 September and 2012 November using the Wide Field Camera \citep[WFCAM;][]{Casali2007} on UKIRT, with a total exposure time of 810s per observation. As observations were not evenly distributed over the survey period, the number of observations compiled into the month stacks ranges from 4 to 214 with an average of 36 for the $J$-band and 50 for the $K$-band stacks. The resulting variety of image depths is accounted for in the estimation of photometric uncertainties \citep[see][]{Elmer2020}. Using these calibrated uncertainties, we find that the S/N of an object with an AB $K$-band magnitude of $\sim 22$ (the faintest magnitude in the sample) in the month snapshots ranges from 2.7 to 15.8, with a mean S/N of 10.1 in the $K$-band and 7.2 in the $J$-band.
    
\begin{figure}
        \centering
        \includegraphics[width=\columnwidth]{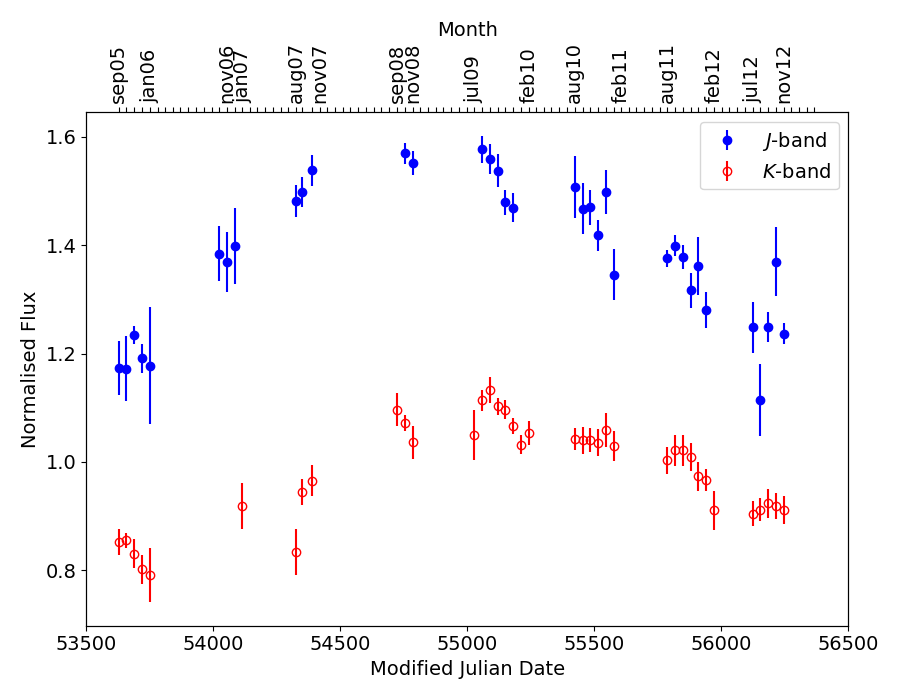}
        \caption{Illustrative $J$- and $K$-band light curves for an object with high $\chi^{2}$ variability. The $K$-band light curve is normalised by the mean flux. The $J$-band data have been offset by 0.4 in normalised flux for clarity.}
        \label{fig:lc}
    \end{figure}

    The resulting $J$ and $K$ band light curves for one of the brighter sources are shown in Figure \ref{fig:lc}.  There is clearly a strong signature of correlated variability in both bands, and even a tantalizing hint of a lag between the two, but when we sought to quantify it using the discrete correlation function (DCF) as described by \cite{Edelson1988}, there proved to be insufficient signal in the individual light curves to robustly quantify their lags. Fortunately, the linear nature of cross-correlations means that we can add together the DCFs from individual light curves, after subtracting their mean values and normalizing them, to calculate their ensemble properties. The only other adjustment that has to be made is to adjust the individual objects for cosmic time dilation, so each light curve is measured in its temporal rest frame; this adjustment had the collateral benefit of smoothing the temporal sampling in the stacked data, eliminating any possibility of residual systematics from the observational sampling. The errors on the DCF are estimated by bootstrapped resampling of the light curves 1000 times.  
    
    If there is a lag between the two light curves, then the peak in the DCF will be offset from zero.  To quantify this effect, following \cite{Edelson1988}, we adopt two measures of the location of the cross-correlation peak.  First, we fit a parabola around the highest single value in the DCF to determine the true location of the peak, $\tau_{\rm peak}$, with an error estimated from the bootstrap analysis as the width of this peak corresponds to the characteristic timescale of variability rather than the error. After experimenting with this fitting, we found that the most robust range to fit this parabola over was -7 to +7 months. Second, we calculate the centroid of all points where $\mathrm{DCF} > 0.5 \times \mathrm{DCF_{max}}$, $\tau_{\rm cent}$, with an error estimated from the standard error on this mean value.  These quantities are complementary in that if there is any structure in the correlation between the bands, $\tau_{\rm peak}$ is a measure the modal lag between the two light curves, while the $\tau_{\rm cent}$ measures the mean lag. 
    
\section{Results}
    \label{sec:Results}
    
\begin{figure}
        \centering
        \includegraphics[width=\columnwidth]{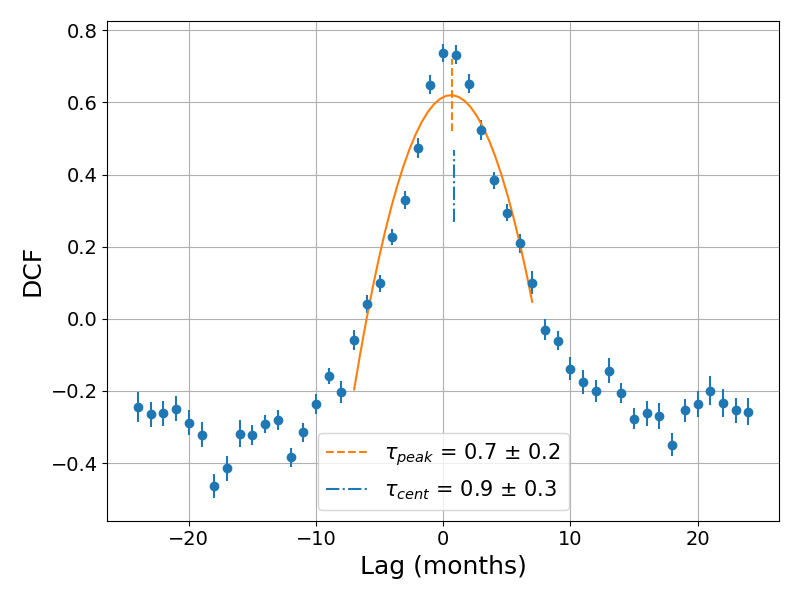}
        \caption{The stacked discrete cross-correlation function between $J$- and $K$-band data for the 94 AGN selected from the UKIDSS UDS.  Two measurements of the characteristic lag are shown (in months). The parabola shows the fit used to calculate $\tau_{\rm peak}$ (shown as an orange dashed line).  The blue dash-dot line shows the centroid shift, $\tau_{\rm cent}$.}
        \label{fig:full_ccf}
    \end{figure}

    We start with an analysis of the full set of 94 light curves, before going on to subdivide the sources by various criteria (see Table \ref{tab:numbers}). The resulting stacked DCF is presented in Figure \ref{fig:full_ccf}. It is apparent that the peak is offset in the positive sense from zero, implying that the $K$-band light lags behind that in the $J$-band, as predicted. Quantitatively, we find $\tau_{\rm peak} = 0.7 \pm 0.2$ months and $\tau_{\rm cent} = 0.9 \pm 0.3$ months, so both metrics are statistically significant at the $3\sigma$ level.  These results illustrate the well-known strength of cross-correlation analysis in detecting signal at, or even below, the binning level, though we note that cosmological time dilation means that the light curves are actually sampled on significantly shorter timescales.  
    
    \begin{table}
        \centering
        \begin{tabular}{| l | c |}
            \hline
            Subset & Number of Variable AGN \\
            \hline
            $\chi^{2} > 100$ & 94 \\
            \hspace{0.5cm}$z \geq 1.2$ & 46 \\
            \hspace{0.5cm}$z < 1.2$ & 41 \\ 
            \hspace{1cm}$-21.1 > M_{K} \geq -24.0$ & 22 \\
            \hspace{1cm}$-24.0 > M_{K} \geq -26.4$ & 19 \\
            \hline

        \end{tabular}
        \caption{The number of variable AGN used to create the discrete correlation functions shown in Figures \ref{fig:full_ccf}-\ref{fig:lum_ccfs}. The total number in the redshift subsets does not add up to 94 as 7 objects were removed due to large uncertainties in their photo-z estimations. }
        \label{tab:numbers}
    \end{table}
    
\begin{figure}
        \centering
        \includegraphics[width=\columnwidth]{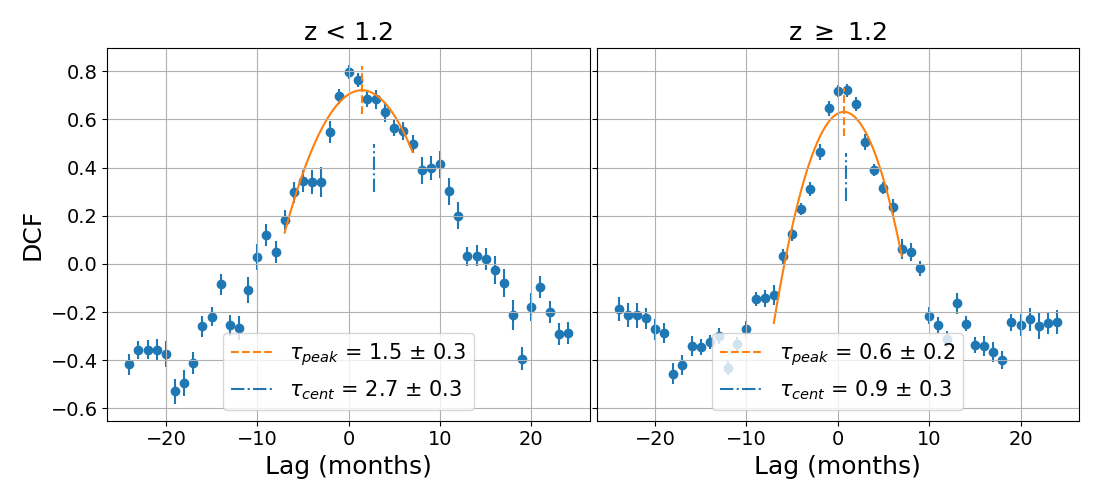}
        \caption{The stacked discrete cross-correlation function, as in Figure~\ref{fig:full_ccf}, but with the AGN divided into two groups based on whether above or below a redshift of $z=1.2$. Two measurements of the characteristic lag are shown (in months).}
        \label{fig:z_ccfs}
    \end{figure}

    One effect that we might expect to dilute the signal in Figure \ref{fig:full_ccf} originates in the wide range of redshifts in the data stacked together.  For the lower redshift sources, while the $J$-band data typically represents optical emission, the $K$-band comes from the rest-frame infrared, providing the signal we are looking for.  However, at higher redshifts, both infrared bands were emitted in the rest-frame optical, for which, as discussed in Section \ref{sec:Intro}, lags are shorter.  To test for this effect, Figure \ref{fig:z_ccfs} shows the stacked DCFs divided into two, with the division at $z = 1.2$ representing the approximate point at which the observed $K$-band corresponds to rest-frame optical light ($<1 \mu$m ).  Indeed, as predicted, the lags are significantly longer in the low-redshift regime where we are comparing rest-frame optical and infrared emission.  It is also notable that the low-redshift DCF peak shows a distinct asymmetry with a longer-lag shoulder, which is further emphasized by the difference between $\tau_{\rm peak}$ and $\tau_{\rm cent}$. There is clearly information here regarding the structure of the emitting region, although a detailed analysis of this transfer function would require more data.  
    
\begin{figure}
        \centering
        \includegraphics[width=\columnwidth]{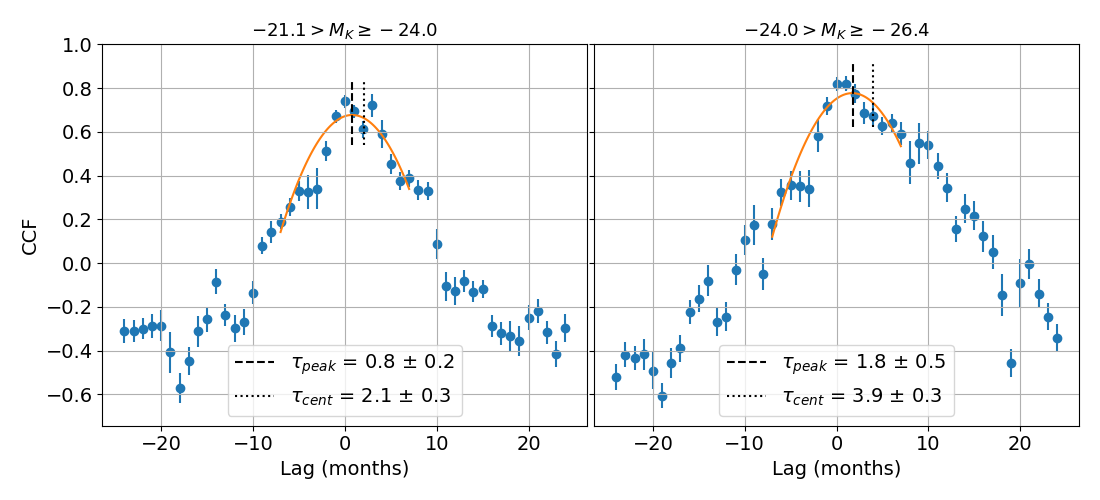}
        \caption{The stacked discrete cross-correlation function, as in Figure~\ref{fig:full_ccf}, but with the low-redshift AGN divided into two groups based on whether their $K$-band absolute magnitude is fainter or brighter than $M_K = -24$. Two measurements of the characteristic lag are shown (in months).}
        \label{fig:lum_ccfs}
    \end{figure}

   One further test that we can apply to this analysis comes from noting that one would generically expect that more luminous sources are physically larger, so the lags might be expected to also be larger for the intrinsically brighter sources.  Figure~\ref{fig:lum_ccfs} shows the DCFs obtained when the lower-redshift sources are separated at a $K$-band absolute magnitude of $M_K = -24$, this divide was chosen as it roughly divides the AGN into equal bins.  As is apparent from this figure, the more luminous sources do, indeed, show significantly longer lags, with the asymmetric shoulder discussed above particularly clear in these objects.
   
    Interestingly, however, if we repeat this analysis using just the higher-redshift sources (where we are comparing emission just in the rest-frame optical), we find a lag of $\sim 1$ month at high luminosities, which is not what we would expect from previous studies of AGN in the rest-frame optical light, which find lags of $\sim 1$ day \citep[e.g.][]{Sergeev2005,Edelson2015}. Therefore, the structure is potentially more complicated than the simple picture we have considered here.  
    
\section{Discussion}
    \label{sec:Discussion}
    
    There is a wealth of information on the structure of AGN to be found in their light curves, and the lags in variability between wave bands.  The near infrared offers a potentially important perspective, as it may probe out to the dusty toroidal region that surrounds an inner accretion disk, but observationally it has proved challenging to study for long enough periods and for a sufficient number sources to learn much about the overall population of AGN.  Here we have shown that the UKIDSS UDS offers a different window on this probe: the period over which it was obtained spans the timescales of interest very nicely, while its depth means that, even within its relatively small field, we can detect orders of magnitude more variable AGN than previous studies of individual targets.  We have also shown that even where there is not sufficient signal to learn much from individual sources, stacking such consistent data together provides a strong ensemble signal. This analysis confirms that there is, indeed, a significantly longer lag between optical and infrared light than between optical bands, and that the length of the lag increases with the luminosity of the source as we might physically expect.
    
    We also see clear indications of structure in the stacked DCFs, indicating that there is still more evidence to be found in data obtained in this way; the fact that this structure does not average away when derived from such stacked data implies that it is a generic feature of these objects rather than a peculiarity of any particular source. Detailed modelling of this signal, and its implications for AGN structure, is beyond the scope of this work. However, we have demonstrated that there is an interesting signal here that is worth modelling and investigating further. Significant progress in this area requires larger areas or deeper data than the UDS supplies, to beat down the remaining noise in this generic signal, and to allow more nuanced subdivision of the data.  Upcoming infrared surveys \citep[e.g. the Euclid deep fields and VEILS:][]{Laureijs2011,Honig2017} will potentially allow us to do exactly this analysis, but only if it is considered in the survey design, for example in how the target region is imaged over time. 
    
\section*{Acknowledgements}
    
    EE is supported by a United Kingdom Science and Technology Facilities Council (STFC) studentship.
    We extend our gratitude to the staff at UKIRT for their tireless efforts in ensuring the success of the UDS project. We also wish to recognise and acknowledge the very significant cultural role and reverence that the summit of Mauna Kea has within the indigenous Hawaiian community. We were most fortunate to have the opportunity to conduct observations from this mountain. This work is based in part on observations from ESO telescopes at the Paranal Observatory (programmes 094.A-0410, 180.A-0776 and 194.A-2003).
    Finally, we would like to thank the referee for their useful and insightful comments which significantly improved the paper. 

%%%%%%%%%%%%%%%%%%%%%%%%%%%%%%%%%%%%%%%%%%%%%%%%%%
\section*{Data Availability}
    
    The raw infrared imaging data used in this paper can be obtained from the WFCAM Science Archive (http://wsa.roe.ac.uk). Further processed data and catalogues will be made available from the UDS web page (https://www.nottingham.ac.uk/astronomy/UDS/). The final DR11 catalogue and photometric redshifts will be presented in Almaini et al. (in preparation) and are available from the authors on request.

% The inclusion of a Data Availability Statement is a requirement for articles published in MNRAS. Data Availability Statements provide a standardised format for readers to understand the availability of data underlying the research results described in the article. The statement may refer to original data generated in the course of the study or to third-party data analysed in the article. The statement should describe and provide means of access, where possible, by linking to the data or providing the required accession numbers for the relevant databases or DOIs.



\bibliographystyle{mnras}
\bibliography{PhD} % if your bibtex file is called example.bib

%\begin{thebibliography}{99}
%\end{thebibliography}

%%%%%%%%%%%%%%%%%%%%%%%%%%%%%%%%%%%%%%%%%%%%%%%%%%

%%%%%%%%%%%%%%%%% APPENDICES %%%%%%%%%%%%%%%%%%%%%

% \appendix

% \section{Some extra material}

% If you want to present additional material which would interrupt the flow of the main paper,
% it can be placed in an Appendix which appears after the list of references.

%%%%%%%%%%%%%%%%%%%%%%%%%%%%%%%%%%%%%%%%%%%%%%%%%%

% Don't change these lines
\bsp	% typesetting comment
\label{lastpage}
\end{document}